\documentstyle[12pt,preprint]{aastex}
\shorttitle{Extrasolar Nebula}
\shortauthors{Kuchner}
\begin{document}

\slugcomment{To appear in ApJ 9/04}

\title{A Minimum-Mass Extrasolar Nebula}
\author{Marc J. Kuchner\altaffilmark{1}}
\affil{Princeton University Observatory \\ Peyton Hall, Princeton, NJ 08544}
\altaffiltext{1}{Hubble Fellow}
\email{mkuchner@astro.princeton.edu}

\begin{abstract}

By analogy with the minimum-mass solar nebula, we construct a surface-density profile
using the orbits of the 26 precise-Doppler planets found in multiple-planet systems:
$\Sigma = 2200 \, \mbox{g} \, \mbox{cm}^{-2} \, (a / {\mbox {1~AU} })^{-\beta}$, where
$a$ is the circumstellar radius and $\beta = 2.0 \pm 0.5$.
The minimum-mass solar nebula ($\beta=1.5$) is consistent with this model, but the uniform-$\alpha$
accretion disk model ($\beta \approx 1$) is not.  In a nebula with $\beta < 2$,
the center of the disk is the likely cradle of planet formation.

\end{abstract}


\section{INTRODUCTION}

Our understanding of the mass profile of
the disk from which the planets of the solar system formed begins with
the minimum mass solar nebula (MMSN).
Imagine adding just enough light elements to each
planet so that all the planets reach solar composition, then smearing the augmented
planets into contiguous nested rings \citep{edge49, kuip56, safr67, alfv70, kusa70, leca73}.
Fitting a power-law to the resulting surface-density distribution
leads to the well-known MMSN prescription,
$\Sigma = \Sigma_0 (a/1 \mbox{AU})^{-3/2}$, where $\Sigma$ is the surface density and $a$
is the circumstellar radius \citep{weid77,haya81}.

The estimates of \citet{weid77} for the masses of the planets augmented to solar composition
suggest that $\Sigma_0 \approx 4200 \, \mbox{g} \, \mbox{cm}^{-2}$.  Using the same table of
solar abundances \citep{came73}, but making some additional assumptions about the
importance of the snow line, \citet{haya81} derived $\Sigma_0 = 1700 \, \mbox{g} \, \mbox{cm}^{-2}$.
Since the snow line's location and role in giant planet formation are
uncertain \citep{sass00}, we prefer the above formula based on \citet{weid77} for
the purposes of this paper.

Elaborate calculations of planet formation \citep[e.g.][]{tril98, ida04} regularly
invoke the MMSN or a $\Sigma \sim a^{-1}$ surface density distribution based
on uniform-$\alpha$ accretion disk models \citep{shak73}.
Such calculations can depend strongly on the assumed surface-density distribution.
For example, the isolation mass produced by runaway planetesimal growth goes as $\Sigma^{3/2}$ \citep{liss87}.  The locations of secular resonances in the nebula are
especially sensitive to the surface-density distribution \citep{leca97, naga00}.

Now, newly discovered extrasolar planets outnumber the solar system planets.
This paper attempts to update our picture of the minimum-mass nebula
taking the extrasolar planets into account;
we derive a surface-density profile analogous to the MMSN using the orbits of the 26 extrasolar planets
found in multiple-planet systems.
Perhaps this new construction---a minimum-mass extrasolar nebula (MMEN)---can serve future
studies of planet formation.

\section{DATA}
\label{sec:data}

The extrasolar planets discovered so far by precise-Doppler methods contain
nine two-planet systems and two three-planet systems.  We obtained the
orbital data for these planets from the website \url{ http://exoplanets.org};
see the review by \citet{marc00} for background on radial-velocity planet-searches.
Table~\ref{tab:planets} summarizes the relevant data on these systems including
measured stellar metallicities from \citet{sant04}.
Of the eleven listed extrasolar planet host stars, eight are G type.

%\clearpage

\begin{table}[h]
\caption{Multiple Planet Systems }
\medskip
\begin{tabular}{llllcc} \hline\hline
Star                  & Spectral & [Fe/H]      &  Semimajor Axes & $\Sigma_0$  & $\beta$ \\
                      & Type     &             &  (AU)           &  ($\mbox{g} \, \mbox{cm}^{-2}$)  & \\
\hline
Sun                   & G2 V   & ~0.0   & 5.20, 9.54, 19.19, 30.07        & 4225 & 1.78 \\
55 Cancri             & G8 V  & ~$0.33 \pm 0.07$ & 0.115, 0.24, 5.9      & 739  & 2.42 \\
$\upsilon$ Andromeda  & F8 V  & ~$0.13 \pm 0.08$  & 0.059, 0.829, 2.53   & 2670 & 1.50 \\
GJ876                 & M4    & ~$0.0^{a}$   & 0.13, 0.21                & 8634 & 2.00 \\
HD 38529              & G4 IV & ~$0.40 \pm 0.06$  & 0.129, 3.71          & 705  & 1.58 \\
HD 82943              & G0    & ~$0.30$  & 0.728, 1.16                   & 8907 & 2.00 \\
HD 169830             & F8 V  & ~$0.21 \pm 0.05$ & 0.81, 3.60            & 2815 & 2.00 \\
HD 12661              & G6 V  & ~$0.36 \pm 0.05$ & 0.82, 2.6             & 3013 & 2.00 \\
HD 168443             & G5 IV & ~$0.06 \pm 0.05$ & 0.295, 2.87           & 3374 & 1.65 \\
HD 74156              & G0    &~$0.16 \pm 0.05$ & 0.294, 3.40            & 1012 & 1.63 \\
47 Ursa Majoris       & G0 V  &~$0.06 \pm 0.03$  & 2.89, 3.73            & 7026 & 2.00 \\
HD 37124              & G4 V  & $-0.38 \pm 0.04$ & 0.54, 2.5             & 1949 & 2.00 \\
$\epsilon$ Eridani    & K7 V   & $-0.13 \pm 0.04$  & 3.3, 40.0            & 715  & 2.00 \\
\hline\hline
\tablenotetext{a}{From \url{http:/exoplanets.org}}
\end{tabular}
\label{tab:planets}
\end{table}

%\clearpage

Besides multiple systems inferred purely from radial velocity observations, the table
also includes $\epsilon$ Eridani. This system contains one planet
inferred from radial-velocity measurements to have a semimajor axis of 3.3~AU and
a second planet inferred from the structure of a circumstellar debris disk
imaged in the submillimeter \citep{grea98, quil02, kuch03}.  \citet{quil02},
estimate that this second planet has a semimajor axis of 40 AU and a mass of $10^{-4}$
 times the mass of $\epsilon$ Eridani (0.8 M${}_{\odot}$).  Since the existence of these
 planets is still debated, we perform our analysis both with and without them in the sample.

Radial velocity surveys are generally biased against detecting low mass
planets at large periods.  Precise-Doppler surveys can presently detect
radial velocity variations with amplitude $\gtrsim 3 \, \mbox{m} \, \mbox{s}^{-1}$.
In terms of the planet's orbital period, $P$, and the ratio of the mass of the
planet to the mass of the star, $\mu$, this detection limit corresponds to
\begin{equation}
\mu \sin{i} \ge 10^{-4} \left({{P} \over {1 \, \mbox{yr}}}\right )^{1/3} \left ({{M} \over {M_{\odot}}} \right)^{-1/3}
\end{equation}
where $i$ is the inclination of the normal to the planet's orbital plane to the line of sight.

%\clearpage

%biasfig.ps
\begin{figure}
\epsscale{0.9}
\plotone{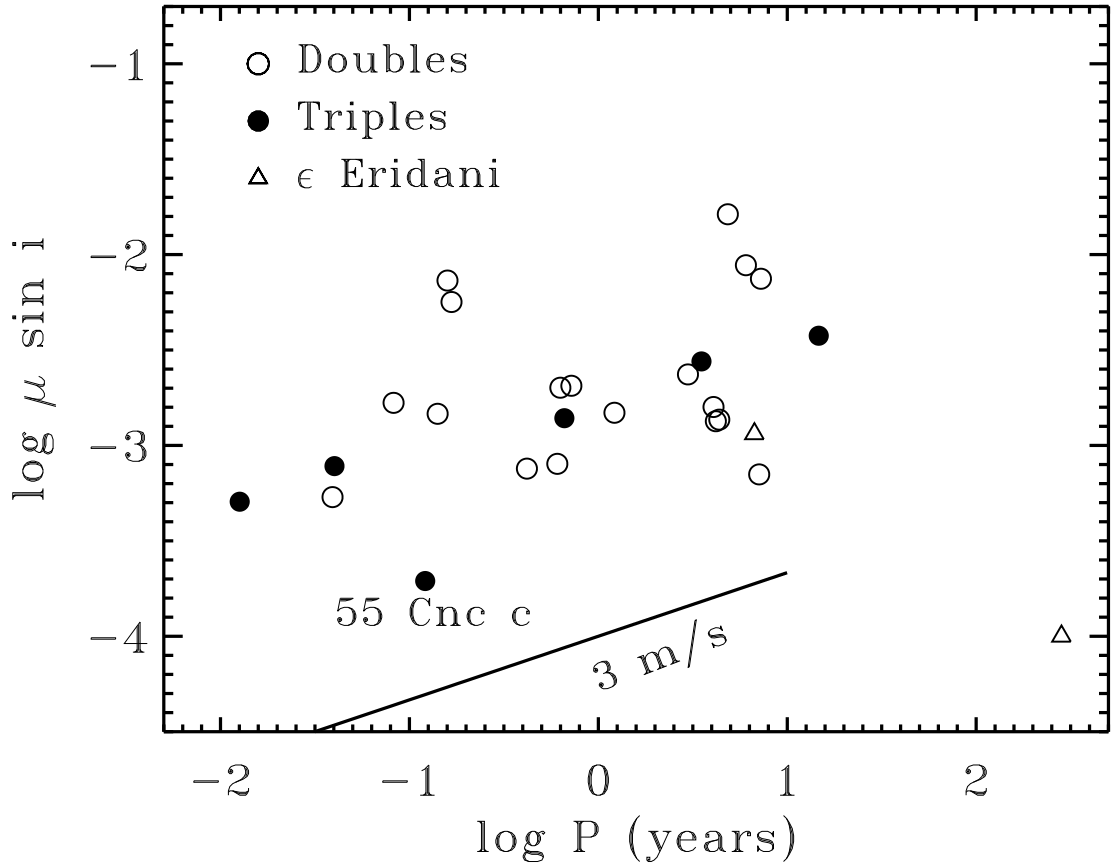}
\caption{Orbital period and $\mu \sin{i}$ for the extrasolar planets in multiple-planet
systems; $\mu$ is the ratio of the mass of the planet to the mass of the star.
The solid line shows the 3 $\mbox{m} \, \mbox{s}^{-1}$ radial-velocity detection limit
for a 1~M${}_{\odot}$ star, comfortably below $\mu \sin{i}$ for most of our sample.
\label{fig:bias}}
\end{figure}

%\clearpage

Figure~\ref{fig:bias} shows the orbital period and $\mu \sin{i}$ of each planet.
Open circles indicate the planets in two-planet systems.  Filled circles
indicate planets in three-planet systems.  The triangles indicate the $\epsilon$ Eridani
planets. The solid line shows the $3 \, \mbox{m} \, \mbox{s}^{-1}$ detection limit,
calculated for a 1 M${}_{\odot}$ star.  For the purpose of Figure~\ref{fig:bias}, we
assume $\sin{i}$ for $\epsilon$ Eridani is $0.5$ based on the inclination of the disk.

This figure suggests that the detection bias in this data set is likely to be small;
all of the precise-Doppler planets in our data set could have been detected out to the
period limit (roughly 10 years), except for 55 Cancri c, which could have been detected
only out to a period of 6 years.  However, other biases might lurk in the process of
reporting and publishing the exoplanet data, and naturally, detecting planets via disk
perturbations suffers from a special set of biases. Our analysis bears repeating when
more data become available.

\section{ANALYSIS}
\label{sec:analysis}

Constructing a MMEN from these data requires associating a surface density with each
extrasolar planet.  To derive these surface densities, we first must estimate
an augmented mass for each planet---the mass each planet would have if it
accreted just enough additional gas with the right chemistry so that it attained solar composition.
We can convert the augmented mass to a surface density based on the spacings of the planets in
multiple-planet systems.

Jupiter has a mass of 318~M${}_{\bigoplus}$ and
roughly 2--4 times solar abundances of C, N and S \citep{gaut01}, suggesting that its augmented mass
should be roughly 1000~M${}_{\bigoplus}$.  Estimates for the augmented masses of Saturn,
Uranus and Neptune are all consistent with this value,
even though their actual masses span a factor of $\sim 20$ \citep{weid77}.
This trend supports a planet formation scenario in which the observed range of
planet masses arises mainly from variations in volatile depletion.

We chose 1000~M${}_{\bigoplus}$ for the augmented masses of most of the extrasolar planets.
If we try to construct the minimum mass solar nebular from the orbits of Jupiter, Saturn,
Uranus and Neptune under this approximation, we get
$\Sigma = 4230 \, \mbox{g} \, \mbox{cm}^{-2} \, (a/{ \mbox{1~AU} })^{-1.8}$, in rough
agreement with the MMSN of \citet{weid77}.
A few planet candidates have measured minimum masses greater than 1000 M${}_{\bigoplus}$:
HD 168443b ($m \sin{i} = 7.73 \, \mbox{M}_{\rm J}$), HD 168443c ($m \sin{i} = 17.2 \, \mbox{M}_{\rm J}$),
HD 74156c ($m \sin{i} = 8.21 \, \mbox{M}_{\rm J}$), $\upsilon$ Andromeda d ($m \sin{i} = 3.75 \, \mbox{M}_{\rm J}$),
and 55 Cancri d ($m \sin{i} = 4.05 \, \mbox{M}_{\rm J}$).  We used the measured minimum masses, $m \sin{i}$,
as the augmented masses for these planets.  Using $m \sin{i}$ for all the augmented masses
would substantially increase the dispersion in the surface densities.

To derive surface densities, we divided each planet's augmented mass by the area of an
appropriate annulus.  Here we harnessed the multiple planet systems; we used
the separation of the planets in log semimajor axis space as the full width of each
annulus.  In other words, we chose the boundaries of the annuli to be the geometric
means of the adjacent semimajor axes.  \citet{weid77} found that using an arithmetic mean or
a geometric mean generated indistinguishable results for the solar system.  However,
the larger separations of the extrasolar planets seem to call for the use of the geometric mean, given the
usual picture of feeding zones for planetary accretion that span widths measured in units of
the local Hill radius \citep[e.g.][]{koku98}.  The extrasolar planets exhibit a much larger
range of orbital eccentricities than the solar planets \citep{marc00}; we do not attempt to take this factor
into account.

%actualmass.ps

%\clearpage

\begin{figure}
\epsscale{0.9}
\plotone{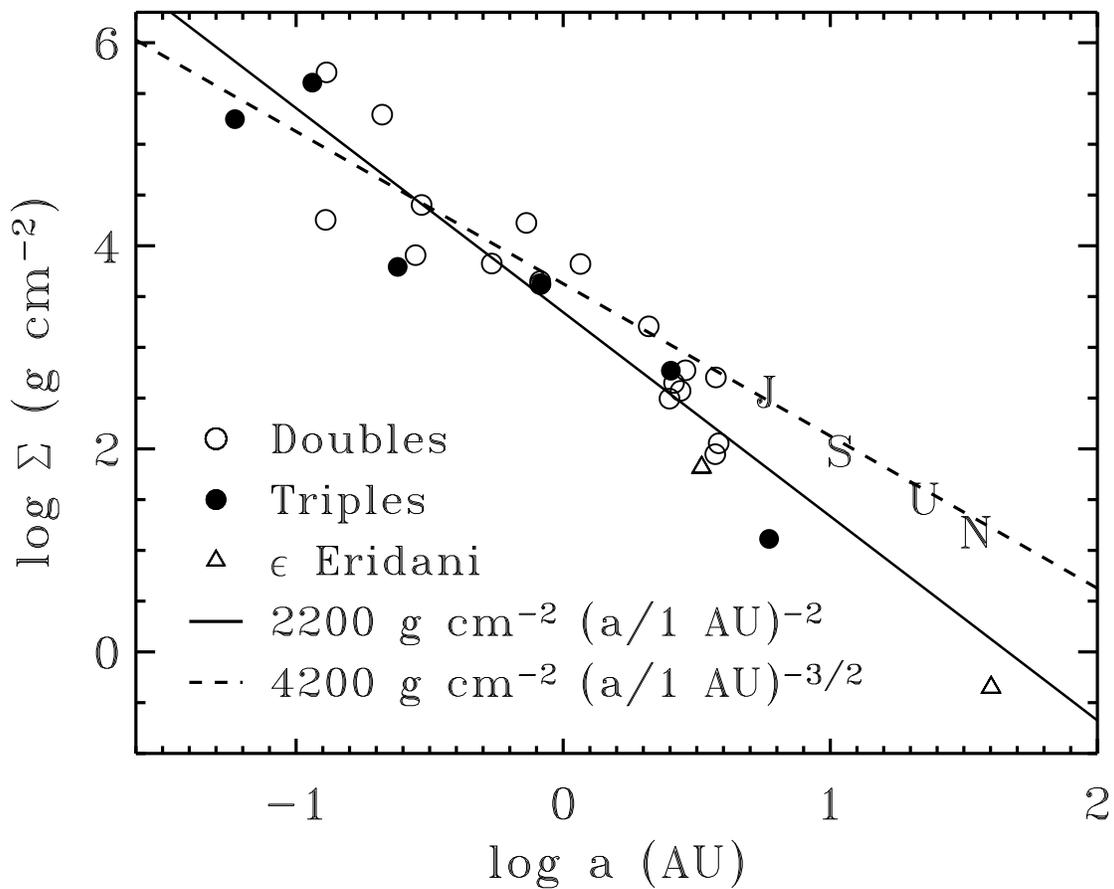}
\caption{Extrasolar planet surface densities. Filled circles are three-planet systems and solid
circles are two-planet systems.  Two triangles indicate the $\epsilon$ Eridani system.  The
best fit power law is $a^{-2}$ (solid line), though $a^{-3/2}$ (dashed line) is also marginally
consistent with the data.
\label{fig:mmen}}
\end{figure}

%\clearpage

Figure~\ref{fig:mmen} shows the surface densities we associate with each planet.  Open circles indicate the
surface densities associated with the planets in two-planet systems.
Filled circles indicate the surface densities associated with the planets in
three-planet systems.  Also plotted are surfaces densities associated with the
$\epsilon$ Eridani planets, marked by triangles.

Table~\ref{tab:planets} reveals some of the subtleties of this procedure.
It lists the parameters $\Sigma_0$ and $\beta$ for minimum-mass nebulae
fitted separately to each system.  The table shows that $\beta=2$ for any two-planet
system where both planets have the same mass; this limitation of the algorithm prevents
us from taking the separately-fitted nebulae too seriously or subdividing the data set
in too much detail.

The table also shows that the metallicities of the host stars range over 0.67 dex.  At first, one might
wonder how this spread in metallicity should affect the augmented masses.  However, it turns out that
for power-law models fit without any attempt to take account of stellar metallicity,
$\Sigma_0$ shows no correlation with stellar metallicity, only scatter.  This lack of a discernible
trend suggests that metallicity corrections are beneath the fidelity of our method.

Table~\ref{tab:power} shows $\Sigma_0$ and $\beta$ for power laws
fit to the data taken as a whole and to subsets of the data.
The power law best fit to the whole data set has the form
$\Sigma = 2200 \, \mbox{g}\,\mbox{cm}^{-2} \, (a/\mbox{1~AU})^{-2.0}$.
This form for the MMEN has a total mass of 8 Jupiter masses in the
region 0.3-30 AU, compared to 31 for the MMSN. This difference may result from the
present inability of radial velocity techniques to measure orbital periods longer
than $\sim 10$ years; we may have only found 1/3 of the mass we will eventually find.

%Changing our prescription for the augmented masses so that the
%Use $m \sin{i}$ for all the augmented masses decreases both the inferred $\Sigma_0$,
%and $\beta$ slightly.  For the two-exoplanet systems, the resulting MMEN is
%$\Sigma = 1580 \, \mbox{g}\,\mbox{cm}^{-2} \, (a/\mbox{1~AU})^{-1.66}$; for the
%three-exoplanet systems, the result is
%$\Sigma = 260 \, \mbox{g}\,\mbox{cm}^{-2} \, (a/\mbox{1~AU})^{-1.92}$.
%This prescription risks correlations between

Figure~\ref{fig:mmen} compares the above form for the MMEN (solid line) with the
standard MMSN, $\Sigma = 4200 \, \mbox{g} \, \mbox{cm}^{-2} \, (a /{1~AU})^{-3/2}$ (dashed line).
Assuming the logarithms of the surface densities have uncertainties of $\pm 0.3$, the uncertainty in the
exponent of the power-law fit to the entire data set is $\beta=2 \pm 0.5$. Given these assumptions,
the standard MMSN is consistent with the exoplanet data.

%\clearpage

\begin{table}[h]
\caption{Best fit surface-density profiles.}
\medskip
\begin{tabular}{lcc} \hline\hline
Data Set                       & $\Sigma_0$ (g cm${}^{-2}$)  & $\beta$  \\
\hline
Two-Planet Systems Only                       & 2990 & 1.87 \\
Three-Planet Systems Only                    & 739   & 2.42 \\
G Star Exoplanets Only                    & 2150  & 1.79 \\
All Exoplanets But $\epsilon$ Eridani & 2500  & 1.87 \\
All Exoplanets    & 2220  & 2.01  \\
Solar System Giants Only        & 4230  & 1.78 \\
Exoplanets + Solar System Giants     & 2490  & 1.87 \\
\hline\hline
\end{tabular}
\label{tab:power}
\end{table}

%\clearpage

\section{DISCUSSION}

\subsection{Other Nebular Models}

By combining the data from several exoplanet systems, we have implicitly
assumed that protoplanetary disks have a common surface-density distribution.
Such a common profile might stem from a steady-state model of the protoplanetary
accretion disk, like the uniform-$\alpha$ accretion disk model \citep{shak73}.

However, the uniform-$\alpha$ accretion disk model ($\beta \approx 1$) is not consistent
with the MMEN power law, and perhaps we should not expect it to be.  This primary
justification for this model is that a uniform $\alpha$ is the simplest assumption one can make about
anomalous disk viscosity.  However, the reason for angular moment transport in disks is
highly debated; models of disk turbulence driven by magneto-rotational instability \citep[e.g.][]{flem02}
suggest that the angular momentum transport is highly non-uniform, and may not even be characterized
by a viscosity.

An alternative steady-state disk model is a massive circumstellar disk bordering on
gravitational instability.   Spiral density waves may
form in these disks that tend to redistribute angular momentum
to maintain Toomre's $Q$ at $\sim 1$ throughout the disk \citep{laug94}.
Or else, planets could conceivably form via direct
gravitational collapse of such disks \citep[e.g.][]{boss01}, depending on the
thermodynamic assumptions \citep{pick00}.
For a disk with uniform $Q$, the surface density profile
is determined by the square root of the local disk temperature, so a variety of disk
thermal models would yield similar surface-density distributions.
These disks are at least an order of magnitude more massive than the minimum mass
nebulae we are considering.  For example, a $Q=1$ disk where the midplane temperature is
$T = 150 \, \mbox{K} \, (a/1~\mbox{AU})^{-3/7}$ \citep{chia97}
has surface density $\Sigma = 88000 \, \mbox{g} \, \mbox{cm}^{-2} (a/1~\mbox{AU})^{-1.93}$.

Of course, the real solar nebula may seldom have been near a steady state; T Tauri stars are
highly variable, and at any given stellar age, one can find stars both with and without accretion
disks in the same cluster \citep[e.g.][]{hais01}.
One-dimensional models of protoplanetary disk accretion more sophisticated than the
uniform-$\alpha$ model \citep{gamm96, mats03} generate time-variable surface density profiles.
Ironically, these models often use the MMSN as a starting point.
In general, a non-steady-state picture of the protoplanetary disk stretches our
interpretation of the MMSN and MMEN; all of the minimum mass need not have been available as free
gas in the disk at the same time!

%sound speed, $c = 9.3 \times 10^4 \, \mbox{cm} \, \mbox{s}^{-1} (a/1~\mbox{AU})^{-3/14}$
%$c \Omega / (\pi G \Sigma)$
%stability means
%$\Sigma < c \Omega / (\pi G) \propto T^{1/2} a^{-3/2}$
%$T \propto a^{-1/2}$ means $\Sigma \propto a^{-1.75}$
%$T \propto a^{-3/7}$  means $\Sigma \propto a^{-1.93}$
%However, recent models of marginally-stable disks have suggested that maintainance
%of $Q \sim 1$ depends on the thermodynamic assumptions \citep{pick00}.
%These simulations are of great importance to scenarios where planets form by
%disk instability \citep[e.g.][]{boss01}; most core accretion scenarios invoke substantially less
%massive disks.

\subsection{A Steeper Surface-Density Distribution?}

Though taken as a whole, the extrasolar planet data agree with the MMSN, Figure~\ref{fig:mmen}
and Table~\ref{tab:power} show that some extrasolar planetary systems themselves do not.
The $\beta$ exponents for the three-exoplanet systems
and for the solar system differ substantially; the minimum-mass nebula constructed from the
three-exoplanet systems has more mass concentrated in the center, like the uniform-Q model mentioned
above.  These three-exoplanet systems are the ones for which our algorithm works the best.

Conventional wisdom holds that giant planets form at 2--5~AU from the star, perhaps at the
snow line, and that the close-in extrasolar planets migrated inward from where they
formed \citep{lin96}.
According to this picture, one might might expect the MMEN to be more centrally peaked than the MMSN.
Perhaps the reason for the steep power
law among the three-exoplanet systems is that these planets are more widely
separated in semimajor axis than planets in the solar system;
with our prescription, increasing planet separations steepens the power law.
The planet-disk interactions that cause the planets to migrate to small $a$ could create the
wider separations; massive planets can clear out the region between them in a protoplanetary
disk via slowly damped spiral density waves \citep{bryd00}, preventing more planets
for growing in these regions.

%Alternatively, inspired by \citet{kuip56} and \citet{leca73}, we might ask: why does the solar
%system lack material in its center compared to extrasolar planetary systems?
%Perhaps most of the metals in the center of the solar system disk were ejected
%in the process of forming the Oort cloud \citep{levi01}. Or perhaps the solar
%system is under-dense in its center because water could not condense there.
%Conceivably, the giant planets in the solar system could even have migrated outward.

\section{INSIDE-OUT PLANET FORMATION}

However, looking at things another way, the steeper nebular surface density distributions
consistent with the MMEN could have dramatic implications for where planets are
formed. Let us briefly indulge this alternative viewpoint.

An annulus of the protoplanetary disk with width
$\Delta r$ has mass $m=2 \pi \Sigma r \Delta r$.  If this annulus is a feeding zone with
$\Delta r \propto r$ and if  $\Sigma \propto r^{-\beta}$, then $m \propto r^{2-\beta}$.
For the minimum-mass solar nebula
($\beta=1.5$) and uniform-$\alpha$ disk model ($\beta \approx 1$), $m$ increases with radius, suggesting that
more mass is available for planet-formation at larger distances from the star.
This tale---an exercise in circular reasoning---has been a standard argument for why giant
planets ought not to form predominantly at the center of the disk.

However, for $\beta > 2$, $m$ decreases with radius, suggesting that the center of
the disk is the most natural site for planet formation.
Perhaps giant planets predominantly form in the center of their disks and migrate outwards.

Indeed, suitable outward migration mechanisms for giant planets exist: Type III
migration \citep{mass03} and planet-planet scattering \citep{thom99,adam03}.
Type III migration begins when the planet grows larger than a
threshold mass sufficient to begin opening a gap in the disk \citep{arty04}.
Some planet-planet interactions may also be required to produce the observed range of extrasolar-planet
orbital eccentricities.

Outward Type III migration will halt at the outer boundary of the disk.
\citet{mats03} have suggested that the disk may develop an edge or a gap at $\sim 5$ AU when
the photoevaporation time scale matches the viscous spreading time scale.  The exact nature
and location of this edge remain
uncertain. However, the disk probably photoevaporates first outside $\sim 5$~AU, providing a
natural stopping place for outward planet migration.

This inside-out planet-formation scenario has several attractive features.
While no direct observational evidence has yet been found for a special structure
at the snow line in protoplanetary disks, abundant new evidence from near infrared
spectroscopy \citep{hart93,muze03} and interferometry \citep{mill01,cola03} implies that T Tauri disks
have dramatic opacity jumps at $\sim 0.1$ AU.
Models of layered accretion disks \citep{gamm96, armi01} suggest that even
while most of the disk has $Q >> 1$, the center of the disk could become gravitationally unstable,
fostering planet formation by disk instability.
Moreover, formation of the planets in the center of the disk could
potentially solve two problems in the core accretion picture:
\begin{trivlist}
\item{\bf Type I Migration of Giant Planet Cores}

Conventional simulations of the semimajor axis distribution of the extrasolar
planets \citep{tril02,armi02,ida04} neglect Type I migration \citep{ward97} on the grounds that no
planets could form at the snow line if this mechanism operated.
Although present calculations of the migration of the small bodies do not accurately
include torques added in the coorbital region \citep{bate03}, these torques are likely to dominate
the net Lindblad torque, not compensate for it, exacerbating the problem.
However, several mechanisms have been suggested that could stop migration
of cores in the center of the disk \citep{lin96, kuch02, terq03, laug03, nels03}.

\item{\bf The Missing Short-Period Pile-up}

\citet{taba02} and \citet{trem03} pointed out that the distribution of planet orbital
periods shorter than $\sim 200$ days is consistent with a power law, and that scenarios where planets
migrate inwards and stop migrating at a special small orbital period should produce a much larger
pile-up of planets at short periods than observed.
\citet{gu03} invoked tidal-driven Roche-lobe overflow to remove this exaggerated pile-up of planets
in 3-day orbits expected from Monte-Carlo migration models.
However, most of the stopping mechanisms cited above are not specific to precisely three-day
periods, and since tidal forces decay as $a^{-6}$, tide-induced Roche-lobe overflow can probably
not help prevent pile-ups at say, a 6-day period.  If planets mostly migrated outwards
when they grew beyond a threshold mass as Type III accretion models suggest,
this missing pile-up problem would not arise.

%\item{3) The time scale for formation of Neptune}

%Because the orbital time scale is so long in the outer solar system,
%oligarchic growth simulations that can barely manage to form
%a 15 M${}_{\bigoplus}$ body at 5 AU in 10 million years can not manage to
%form such a body at 30 AU in the lifetime of the protoplanetary disk \citep{thom03}.
%Perhaps Neptune formed closer in where the dynamical time is short \citep{thom99}.
\end{trivlist}

Two effects make the very center of the disk (0.05-0.1 AU) problematic for core acretion:
heating of the planet by tidal interaction with the star \citep{gu03}
and the large critical core radius for gas accretion at high nebular temperatures \citep{bode00}.
However, the region at $\sim 0.2$~AU could be fertile ground for planet formation by either
core accretion or disk instability.
The disk temperature there is potentially low enough that the critial core radius shrinks to where it
is consistent with the inferred core mass of Jupiter and Saturn \citep{papa99, ikom01}
particularly if this region is shadowed by a puffed-up inner wall of the disk \citep{natt01,dull01}.
Conceivably, the observed extrasolar planets
all could have formed at 0.1--0.2~AU and migrated outwards to their current orbits.
The solar system giants might also have formed via this process, though this possibility
raises the question of the origin of the abundant low-temperature condensates in the envelopes
of these planets \citep[e.g.][]{gaut01}, a question still inadequately answered by any
planet-formation model.

\section{CONCLUSIONS}

By analogy with the minimum mass solar nebula, we
constructed the surface density distribution for a minimum-mass extrasolar nebula
based on the orbits of the extrasolar planets in two- and three-planet systems:
$\Sigma \approx 2200 \, \mbox{g} \, \mbox{cm}^{-2} \, (a/\mbox{1~AU})^{-2}$.
The standard MMSN is consistent with the exoplanet data taken as a whole, though the
uniform-$\alpha$ accretion disk \citep{shak73} is not.

In general, though the MMSN is consistent with the MMEN, our analysis
illustrates that the logic that led to the MMSN supports a wide range of disk
surface-density profiles, none of which can serve as a complete picture.
The MMSN was based on Laplace's concept of the solar nebula as a smooth disk that
broke up into rings that condensed into planets.  In contrast, true protoplanetary
disks are likely actively accreting and time variable.

Our analysis led us to consider the ramifications of alternative surface-density
distributions.  We pointed out that in the centrally-concentrated nebular model
suggested by the three-exoplanet systems, the center of the disk is the preferred site for
planet formation.  The extrasolar giant planets---and possibly the giant planets
of the solar-system---could have formed at $0.1-0.2$~AU and migrated outwards to
their present orbits via Type III migration and planet-planet scattering.

%In general, we should recognize that some theoretical predictions involve
%interpolating or extrapolating the minimum mass solar nebula or some other power law.
%The best-fit power law MMEN ($\beta=2$) predicts a density in the
%Kuiper Belt 6 times lower than the MMSN.
Our simple model is not intended to replace the MMSN, but to illuminate its shortcomings and
to expose our solar-centrism.  We must not remain chained to $\beta=1.5$; minimum mass
can sometimes mean maximum prejudice.

\acknowledgments

Thanks to Ed Thommes, Jack Lissauer, Norm Murray, Andrew Youdin, Mike Lecar, Pawel Artymowicz,
Scott Tremaine, David Stevenson, and an anonymous referee for comments.

M.J.K. acknowledges the support of the Hubble Fellowship Program of the Space Telescope
Science Institute and the Kavli Institute of Theoretical Physics.
This research was supported in part by the National Science Foundation under
Grant No. PHY99-0794


\begin{thebibliography}{}

\bibitem[Adams \& Laughlin(2003)]{adam03}Adams, F.~C. \& Laughlin, G. 2003, Icarus, 163, 290

\bibitem[Alfv\'en \& Arrhenius(1970)]{alfv70}Alfv\'en, H. \& Arrhenius, G. 1970, Ap\&SS, 8, 338

\bibitem[Armitage et al.(2001)]{armi01}Armitage, P.~J., Livio, M., \& Pringle, J.~E. 2001, \mnras, 324, 705

\bibitem[Armitage et al.(2002)]{armi02}Armitage, P.~J., Livio, M., Lubow, S.~H., \& Pringle, J.~E. 2002, \mnras, 334, 248

\bibitem[Arymowicz(2004)]{arty04}Artymowicz, P. 2004, talk at the Kavli Institute for Theoretical Physics, \url{http://online.itp.ucsb.edu/online/planetf\_c04/artymowicz/}

\bibitem[Bate et al.(2003)]{bate03}Bate, M.~R., Lubow, S.~H., Ogilvie, G.~I. \& Miller, K.~A. 2003, \mnras, 341, 213

\bibitem[Bodenheimer et al.(2000)]{bode00}Bodenheimer, P., Hubickyj, O., \& Lissauer, J. 2000, Icarus, 143, 2

\bibitem[Boss(2001)]{boss01}Boss, A. 2001, \apj, 563, 367

\bibitem[Bryden et al.(2000)]{bryd00}Bryden, G. Rozyczka, M., Lin, D.~N.~C. \& Bodenheimer, P. 2000, \apj, 540, 1091

\bibitem[Cameron(1973)]{came73}Cameron, A.~G.~W. 1973, Space Sci. Rev., 15, 121

\bibitem[Chiang \& Goldreich(1997)]{chia97}Chiang, E. \& Goldreich, P. 1997, \apj, 490, 368

\bibitem[Colavita et al.(2003)]{cola03}Colavita, M. et al. 2003, \apj, 592, L83

\bibitem[Dullemond et al.(2001)]{dull01}Dullemond, C.~P., Dominik, C. \& Natta, A. 2001, \apj, 560, 957

\bibitem[Edgeworth(1949)]{edge49}Edgeworth, K.~E. 1949, \mnras, 109, 600

\bibitem[Fleming \& Stone(2002)]{flem02}Fleming, T. \& Stone, J. M. 2002, \apj, 585, 908

\bibitem[Gammie(1996)]{gamm96}Gammie, C. F. 1996, \apj, 462, 725

\bibitem[Gautier et al.(2001)]{gaut01}Gautier, D., Hersant, F., Mousis, O. \& Lunine, J.~I. 2001, \apj, 550, L227

\bibitem[Greaves et al.(1998)]{grea98}Greaves J.~S. et al. 1998, \apj, 506, L133

\bibitem[Gu et al.(2003)]{gu03}Gu, P.-G., Lin, D.~N.~C., \& Bodenheimer, P.~H. 2003, \apj, 588, 509

\bibitem[Haisch et al.(2001)]{hais01}Haisch, K.~E.,, Lada, E.~A. \& Lada, C.~J.2001, \apj, 553, L153

\bibitem[Hartmann et al.(1993)]{hart93}Hartmann, L., Kenyon, S.~J., Calvet, N. 1993, \apj, 407, 219

\bibitem[Hayashi(1981)]{haya81}Hayashi, C. 1981, Prog. Theor. Phys. Suppl., 70, 35

\bibitem[Ida \& Lin(2004)]{ida04}Ida, S. \& Lin, D.~N.~C. 2004, \apj, 604, 388

\bibitem[Ikoma et al.(2001)]{ikom01}Ikoma, M., Emori, H. \& Nakazawa, K. 2001, \apj, 553, 999

\bibitem[Kokubo \& Ida(1998)]{koku98}Kokubo, E. \& Ida, S. 1998, Icarus, 131, 171

\bibitem[Kuchner \& Holman(2003)]{kuch03}Kuchner, M.~J. \& Holman, M.~J. 2003, \apj, 588, 1110

\bibitem[Kuchner \& Lecar(2002)]{kuch02}Kuchner, M.~J. \& Lecar, M. 2002, \apj, 574, L87

\bibitem[Kuiper(1956)]{kuip56}Kuiper, G. P. 1956, J. Roy. Astron. Soc. Canada, 50, 158

\bibitem[Kusaka et al.(1970)]{kusa70}Kusaka, T., Nakano, T. \& Hayashi, C. 1970, Prog. Theor. Phys, 44, 1580

\bibitem[Laughlin et al.(2003)]{laug03}Laughlin, G., Steinacker, A. \& Adams, F. 2003, submitted to the \apj, (astro-ph/0308406)

\bibitem[Laughlin(1994)]{laug94}Laughlin, G., \& Bodenheimer, P. 1994, \apj, 436, 335

\bibitem[Lecar \& Franklin(1973)]{leca73}Lecar, M. \& Franklin. F. 1973, Icarus, 20, 422

\bibitem[Lecar \& Franklin(1997)]{leca97}Lecar, M. \& Franklin, F. 1997, Icarus, 129, 134

\bibitem[Levison et al.(2001)]{levi01}Levison, H.~F., Dones, L., \& Duncan, M. J. 2001, \aj, 121, 2253

\bibitem[Lissauer(1987)]{liss87}Lissauer, J. J. 1987, Icarus, 69, 249

\bibitem[Lin et al.(1996)]{lin96}Lin, D.~N.~C., Bodenheimer, P. \& Richardson, D.~C. 1996, Nature, 380, 606

\bibitem[Millan-Gabet et al.(2001)]{mill01}Millan-Gabet, R., Schloerb, F.~P., \& Traub, W.~A. 2001, \apj, 546, 354

\bibitem[Marcy \& Butler(2000)]{marc00}Marcy, G.~W. \& Butler, R.~P. 2000, \pasp, 12, 137

\bibitem[Masset \& Papaloizou(2003)]{mass03}Masset, F. S. \& Papaloizou, J.~C.~B. 2003, \apj, 588, 494

\bibitem[Matsuyama et al.(2003)]{mats03}Matsuyma, I., Johnstone, D. \& Murray, N. 2003, \apj, 585, L143

\bibitem[Muzerolle et al.(2003)]{muze03}Muzerolle, J., Hillenbrand, L. \& Calvet, N. 2003, 592, 266

\bibitem[Nagasawa \& Ida(2000)]{naga00}Nagasawa, M. \& Ida, S. 2000, \aj, 120, 3311

\bibitem[Natta et al.(2001)]{natt01}Natta, A., Prusti, T., Neri, R., Wooden, D., Grini, V.~P., \& Mannings, V. 2001,  \aap, 371, 186

\bibitem[Nelson \& Papaloizou(2003)]{nels03}Nelson, R.~P. \& Papaloizou, J.~C.~B. 2003, submitted to \mnras, (astro-ph/0308360).

\bibitem[Papaloizou \& Terquem(1999)]{papa99}Papaloizou, J.~C. \& Terquem, C. 1999, \apj, 521, 823

\bibitem[Pickett et al.(2000)]{pick00}Pickett, B.~K., Cassen, P., Durisen, R.~H., \& Link, R. 2000, \apj, 529, 1034

\bibitem[Quillen \& Thorndike(2002)]{quil02}Quillen, A. C. \& Thorndike, S. 2002, \apj, 578, L149

\bibitem[Safronnov(1967)]{safr67}Safronov, V.~S. 1967, Soviet Astronomy, 10, 650

\bibitem[Santos et al.(2004)]{sant04}Santos, N.~C., Israelian, G. \& Mayor, M. 2004, \aap, 415, 1153

\bibitem[Sasselov \& Lecar(2000)]{sass00}Sasselov, D.~D. \& Lecar, M. 2000, \apj, 528, 995

\bibitem[Shakura \& Sunyaev(1973)]{shak73}Shakura, N.~I. \& Sunyaev, R.~A. 1973, \aap. 24. 337

%\bibitem[Sylvester et al.(1996)]{sylv96}Sylvester, R.~J., Skinner, C.~J., Barlow, M.~J., \& Mannings, V. 1996, \mnras, 279, 915

\bibitem[Tabachnik \& Tremaine(2002)]{taba02}Tabachnik, S. \& Tremaine, S. 2002, \mnras, 335, 151

\bibitem[Terquem(2003)]{terq03}Terquem, C.~E.~J.~M.~L.~J. 2003, \mnras, 341, 1157

\bibitem[Thommes et al.(1999)]{thom99}Thommes, E.~W., Duncan, M.~J., \& Levison, H.F. 1999, Nature, 402, 635

%\bibitem[Thommes et al.(2003)]{thom03}Thommes, E.~W., Duncan, M.~J., \& Levison, H.~F. 2003, Icarus, 161, 431

\bibitem[Tremaine \& Zakamska(2003)]{trem03}Tremaine, S. \& Zakamska, N.~L. 2002,  proceedings of ``The Search for Other Worlds'' conference, October, 2003, (astro-ph/0312045)

\bibitem[Trilling et al.(1998)]{tril98}Trilling, D.~E. et al. 1998, \apj, 500, 428

\bibitem[Trilling et al.(2002)]{tril02}Trilling, D.~E., Lunine, J.~I., \& Benz, W. 2002, \aap, 394, 241

\bibitem[Ward(1997)]{ward97}Ward, W.~R. 1997, \apj, 482, L211

\bibitem[Weidenschilling(1977)]{weid77}Weidenschilling, S.~J. 1977, Astropys. Space Sci., 51, 153



\end{thebibliography}
\end{document}